\documentclass{ws-p8-50x6-00}

\def\bea{\begin{eqnarray}}
\def\beann{\begin{eqnarray*}}
\def\eea{\end{eqnarray}}
\def\eeann{\end{eqnarray*}}
\begin{document}

\title{ Quark interchange effects in the KN interaction}
\author{D. Hadjimichef}

\address{Instituto de F\'{\i}sica e Matem\'{a}tica, UFPel, Caixa Postal 354\\
96010-900, Pelotas, RS.}

\author{J. Haidenbauer}

\address{ Forschungszentrum J\"ulich, Institut f\"ur Kernphysik, 
D-52425 J\"ulich, Germany} 

\author {G. Krein}
\address{ Instituto de F\'{\i}sica Te\'{o}rica, Universidade Estadual 
Paulista  \\  Rua Pamplona, 145 - 01405-900 S\~{a}o Paulo, SP, Brazil}

\maketitle

\abstracts{
 We study the short range repulsion in the KN system due
to quark-gluon exchange. Phase shifts for spin-spin, color Coloumb
and spin-orbit interactions are presented.}


It has been argued \cite{kn} that a more suitable investigation of the short
range part of the hadronic repulsion could be made in the $KN$ system. In the 
constituent quark model the $K^{+}=u\,\bar{s}$ and $K^{-}=\,s\,\bar{u}$ 
while the nucleon is a $uud$ or $ddu$ state so, in the non-relativistic limit,
simple quark exchange mechanism can be applied to the $K^{+}N$ system. 
Many groups have studied
this system in the S-wave \cite{kn-many}. An extension in
order to include higher partial waves, such as the P-waves, is the 
subject of our research.

The method we employ in order to introduce the quark-gluon degrees of
freedom is known as the Fock-Tani formalism \cite{ft}.  
The central idea in the Fock-Tani method is the change of representation
concept. The operators of the composite particles are redescribed by ideal
operators which obey canonical (anti)commutation relations. These ideal
operators act on an enlarged Fock space which is a graded direct product of
the original Fock space and an ``ideal state space'' \cite{ft}. 
The ideal operators correspond to particles with the same quantum numbers as 
the composite ones of the system. A change of representation is implemented 
by means of a
unitary transformation, which transforms the single-composite states into
single-ideal states. When the unitary transformation is applied to the
microscopic quark-quark Hamiltonian one obtains the effective interaction.
The meson-baryon potential can be obtained applying in a standard way the 
Fock-Tani transformed operators to the microscopic Hamiltonian \cite{dimi0}:
\begin{equation}
V_{{\rm meson-baryon}}=\sum_{i=1}^4\;V_i(\alpha \beta ;\delta \gamma )\;{{%
m_\alpha ^{\dagger }\,}b_\beta ^{\dagger }\,m_\gamma \,b_\delta }
\label{vkn_1}
\end{equation}
and 
\begin{eqnarray*}
V_1{(\alpha \beta ;\delta \gamma )} &=&-3V_{qq}(\mu \nu ;\sigma \rho )\,\Phi
_\alpha ^{*\mu \nu _2}\Psi _\beta ^{*\nu \mu _2\mu _3}\Phi _\gamma ^{\rho
\nu _2}\Psi _\delta ^{\sigma \mu _2\mu _3}\;\   \label{vkn_2} \\
V{_2{(\alpha \beta ;\delta \gamma )}} &=&{-3V_{q\overline{q}}(\mu \nu
;\sigma \rho )\,\Phi _\alpha ^{*\mu _1\nu }\Psi _\beta ^{*\mu \mu _2\mu
_3}\Phi _\gamma ^{\sigma \rho }\Psi _\delta ^{\mu _1\mu _2\mu _3}}  \nonumber
\\
V{_3{(\alpha \beta ;\delta \gamma )}} &=&{-3V_{qq}(\mu \nu ;\sigma \rho
)\,\Phi _\alpha ^{*\mu \nu _2}\Psi _\beta ^{*\mu _1\nu \mu _3}\Phi _\gamma
^{\mu _1\nu _2}\Psi _\delta ^{\sigma \rho \mu _3}}  \nonumber \\
V{_4{(\alpha \beta ;\delta \gamma )}} &=&{-6V_{q\overline{q}}(\mu \nu
;\sigma \rho )\,\Phi _\alpha ^{*\nu _1\nu }\Psi _\beta ^{*\mu _1\mu \mu
_3}\Phi _\gamma ^{\mu _1\rho }\Psi _\delta ^{\nu _1\sigma \mu _3}.} 
\nonumber
\end{eqnarray*}
The general potential (\ref{vkn_1}) can be specialized to study the KN system
\cite{dimi}.
Details of this calculation, for the higher partial waves, will be shown 
elsewhere \cite{dimi2}. In the present calculation for the S and P waves 
we will use the microscopic quark-quark potential and quark-antiquark
potential written as 
\begin{eqnarray*}
V_{\mbox{\rm $qq$ or $q\overline{q}$ }}^{\mathrm{OGEP}} &=&
4\pi \alpha _s\left[\,\,
\frac {1}{q^2}
-\frac 2{3m_1m_2}{{\bf S}_{i}}\cdot {{\bf S}_{j}} 
+
\frac {i}{q^2}
\left\{ 
\frac{ 
{\bf S}_{1}
\cdot (
{\bf q}
\times 
{\bf p}_{1})}{2m_1^2}
-\frac{ 
{\bf S}_{1}\cdot ({\bf q}\times {\bf p}_2)}{m_1m_2} \right.  \right.
\nonumber \\
&& \left.\left. -
\frac{
{\bf S}_2
\cdot (
{\bf q}
\times 
{\bf p}_2)}{2m_1^2}
+\frac{ {\bf S}_2\cdot ({\bf q}\times {\bf p}_1)}{m_1m_2}
\right\} \,\,
\right] 
\end{eqnarray*}
The first term is the color Coulomb interaction, the second term is the 
spin-spin interaction and last one is the spin-orbit interaction.
A scattering amplitude can be obtained from the Fock-Tani formalism  such that
the KN OGEP (One-Gluon Exchange Potential) can give rise, in the Born 
approximation, to an amplitude $h^{KN}_{fi}$
presented in ref. \cite{barnes}. The phase shifts can be calculated by
\beann
\delta _{l}^{{\rm KN}}=-\frac{2\pi ^{2}P_{cm}E_{K}\,E_{N}}{\ E_{K}+E_{N}}%
\int_{-1}^{1}d\mu \;h_{fi}^{{\rm KN}}P_{l}(\mu )
\eeann
where $P_{l}(\mu )$ is the Legendre polynomial of  order $l$ and 
$E_{i}=\sqrt{P^2_{cm} +m^2_i}$.

In Figs. 1 and 2 phase shifts are presented for the following partial
waves: $S_{01}$, $S_{11}$, $P_{01}$, $P_{11}$, $P_{03}$, $P_{13}$ 
(where the notation $L_{I,2J}$ is used).

\underline{{\bf Acknowledgements}}

 Financial support for this work was provided in
part by the international exchange program DLR (Germany, BRA W0B 2F)
- CNPq (Brazil, 910133/94-8). One of the authors (DH) is also supported by 
FAPERGS.


\begin{figure}[b]
\epsfxsize=15pc 
\par
\begin{center}
\epsfbox{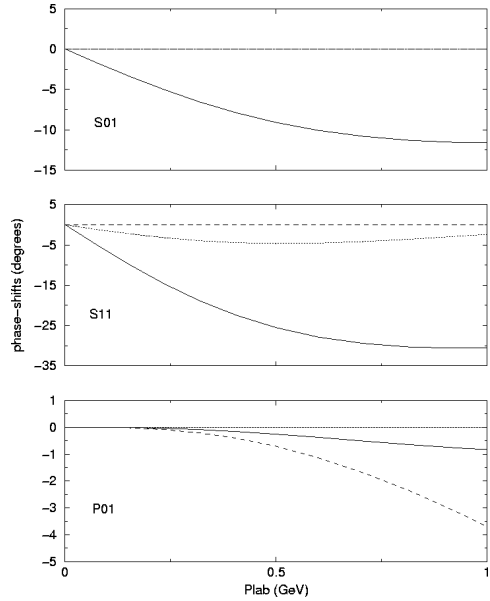}
\end{center}
\caption{$KN$ phase-shifts: spin-spin solid line, Coulomb dotted line, 
spin-orbit dashed line. }
\end{figure}

\begin{figure}[t]
\epsfxsize=20pc 
\par
\begin{center}
\epsfbox{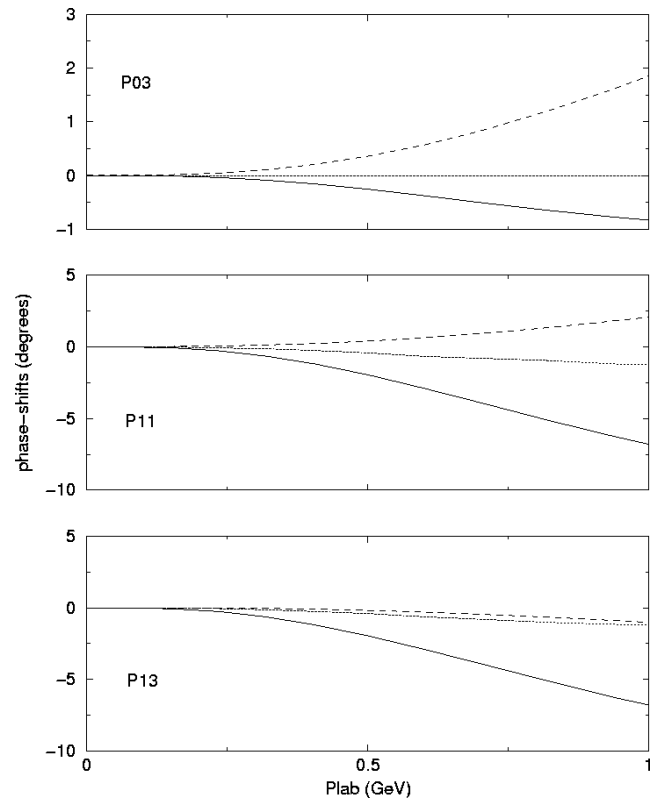}
\end{center}
\caption{$KN$ phase-shifts: spin-spin solid line, Coulomb dotted line, 
spin-orbit dashed line. }
\end{figure}


\begin{thebibliography}{9}


\bibitem{kn}  R. B\"{u}ttgen, K. Holinde, D. Lohse, A. M\"{u}ller-Groeling,
J. Speth, P. Wyborny, {\sl Z. Phys}. {\bf C46} - Particles and Fields, S167
(1990).

\bibitem{kn-many}  J. Bender, H. G. Dosch, H. J. Pirner and H. G. Kruse, 
{\sl Nucl. Phys.} {\bf A414}, 359 (1984); D. Mukhopadyay and H. J. Pirner, 
{\sl Nucl. Phys.} {\bf A442}, 605 (1985); R. K. Campbell and D. Robson, 
{\sl Phys. Rev} {\bf D36}, 2682 (1987).

\bibitem{ft}  D. Hadjimichef, G. Krein, S. Szpigel, J. S. da Veiga, {\sl %
Phys. Lett. }{\bf B367},{\sl \ }317 (1996); 
D. Hadjimichef, G. Krein, S. Szpigel, J. S. da Veiga, 
{\sl Ann. of Phys}. {\bf 268}, 105 (1998).

\bibitem{dimi0}  D. Hadjimichef, {\tt nucl-th/9912022}. 
\bibitem{dimi} D. Hadjimichef, {\tt hep-ph/0006330}. 
\bibitem{dimi2} D. Hadjimichef, J. Haidenbauer, and G. Krein, in
preparation. 
\bibitem{barnes}  T. Barnes, E. S. Swanson, {\sl Phys. Rev}.{\bf \ C49},1166
(1994).
\end{thebibliography}
\end{document}